\documentclass[preprint]{vgtc}               

\graphicspath{{figures/}{pictures/}{images/}{./}} 

\usepackage{enumitem}

\usepackage{times}                     

\usepackage{tabu}                      
\usepackage{booktabs}                  
\usepackage{lipsum}                    
\usepackage{mwe}                       
\usepackage{tabularx} 

\usepackage{mathptmx}                  

\onlineid{TODO}

\vgtccategory{Research}

\vgtcinsertpkg

\title{Towards Deeper Understanding of Natural User Interactions in Virtual Reality Based Assembly Tasks}

\author{
    Ryan Ghamandi\thanks{e-mail: ryanghamandi1@gmail.com}\\
    \scriptsize University of Central Florida
    \and
    Yahya Hmaiti\thanks{e-mail: Yohan.Hmaiti@ucf.edu}\\
    \scriptsize University of Central Florida
    \and
    Mykola Maslych\thanks{e-mail: mykola.maslych@ucf.edu}\\
    \scriptsize University of Central Florida
    \and
    Ravi Kiran Kattoju\thanks{e-mail: kattoju.ravikiran@gmail.com}\\
    \scriptsize University of Central Florida
    \and
    Joseph J. LaViola Jr.\thanks{e-mail: jlaviola@ucf.edu}\\
    \scriptsize University of Central Florida
}

\teaser{
  \centering
  \includegraphics[width=.8\linewidth]{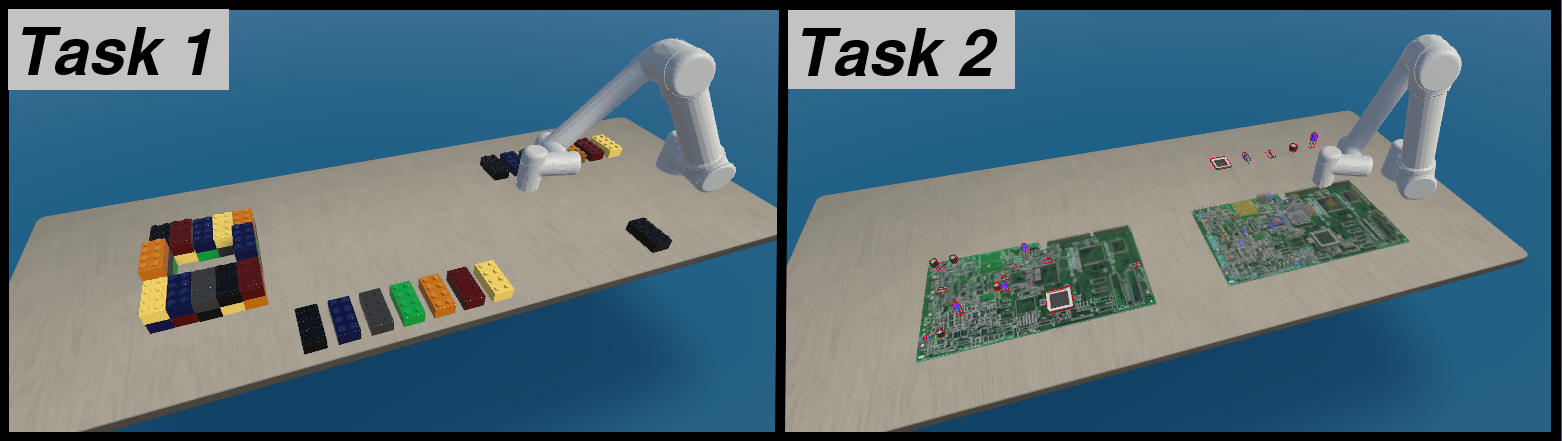}
  \caption{Task setups for LEGO and PCB tasks during completion.}
  \label{fig:teaser}
}

\abstract{
We explore natural user interactions using a virtual reality simulation of a robot arm for assembly tasks. Using a Wizard-of-Oz study, participants completed collaborative LEGO and instructive PCB assembly tasks, with the robot responding under experimenter control. We collected voice, hand tracking, and gaze data from users. Statistical analyses revealed that instructive and collaborative scenarios elicit distinct behaviors and adopted strategies, particularly as tasks progress. Users tended to use "put-that-there" language in spatially ambiguous contexts and more descriptive instructions in spatially clear ones. Our contributions include the identification of natural interaction strategies through analyses of collected data, as well as the supporting dataset, to guide the understanding and design of natural multimodal user interfaces for instructive interaction with systems in virtual reality.

} 


\begin{document}


\firstsection{Introduction}

\maketitle

Interaction with systems allows individuals to tell a system what to do or how to perform~\cite{van_der_veer_interaction_1985}. The medium through which users communicate with a system is the user interface, to which users provide inputs. Natural User Interfaces (NUIs)~\cite{petersen2009continuous} take natural actions as input to a system, with Natural Multimodal User Interfaces (NMUIs) accepting multiple inputs at a time such as voice and gestures. These serve to significantly improve user experiences by supporting more intuitive, engaging, and effective ways to interact~\cite{laviola2014multimodal, oviatt2000designing}. Work has gone towards creating NMUIs for industrial applications such as modeling and assembly, starting with early works that employ speech-based NUIs~\cite{selfridge1986natural, bolt1980put}. From there, they have evolved into NMUIs for scenarios such as virtual reality (VR)~\cite{geiger2020natural, zachmann2001natural, ikonomov2004virtual}, augmented reality~\cite{siltanen2007multimodal, chu2023augmented, wang2016multi}, and CAD modeling using gestural inputs for manual manipulation~\cite{fechter2022comparative, fiorentino2013design}. 

These interfaces enable users to instruct a system in the same way they directly interact with the environment. However, achieving this natural interaction poses challenges for delivering instructions effectively. Systems tend to require users to provide natural inputs in specific patterns. Without the necessary training and/or domain knowledge, this can lead to misunderstandings between the system and user due to misaligned inputs~\cite{10.1145/221296.221302}. In the assembly domain, these interfaces support applications such as rapid prototyping, training, and robotic surgery, enabling seamless operation through natural user instruction \cite{castillo2009virtual, ahmad2015rapid, reddy2023advancements, iftikhar2024artificial}. Recent work in this domain has increased understanding of how users provide inputs to systems, but still have key limitations. Interactions are fueled by prior instruction provided to the user on how to interact with the system, leading to outcomes that represent taught behaviors rather than natural, intuitive interactions. Additionally, current research does not account for intuitive multimodal interactions, particularly in VR, where interaction modalities may vary from real-world tasks despite similar expected outcomes~\cite{joyner2021comparison, rose2000training}. A lack of investigation into capturing sequences of raw, unstructured, and intuitive multimodal input highlights the need for a more nuanced understanding of user behavior in relation to the development of robust and adaptive NMUIs.

To address this gap, we investigate how users naturally instruct a system to perform desired actions, we conducted a single-factor elicitation study to explore how users naturally interact with a virtual robot arm in VR assembly tasks. Participants completed two types of ecologically valid tasks with a Wizard-of-Oz controlled virtual robot arm. In the first task, \textit{collaborative LEGO assembly}, both the participant and the robot manipulated LEGO to recreate a reference model, where LEGO acted as relative spatial anchors that made referencing easy. In the second task, an \textit{instructive Printed Circuit Board (PCB) assembly}, participants instructed the robot to place PCB components on a blank board based on a reference design where the locations for placing objects would be spatially ambiguous to describe. VR was used as a medium to aid in data collection.

Our analysis revealed that users predominantly adopted Bolt's "put-that-there" method~\cite{bolt1980put}, with different language structures of the same commands being delivered. Instruction delivery also varied sequentially, aligning with findings from Oviatt et al.~\cite{oviatt2000designing, 10.1145/319382.319398}. In addition to these findings, we found that descriptive commands and spatially vague instructions accompanied by spatial cues were common in the LEGO and PCB task respectively. Also, participants used more unimodal commands during the collaborative task, more trimodal commands during the instructive task, and bimodal commands in both. Finally, the method by which users would reference locations in the environment for object placement had both spatial and temporal dependencies which was rooted in the availability of concrete spatial anchors.
We list our main contributions as follows: \textbf{(1)} \textit{Key observations of tendencies and statistical analyses of user interactions with a simulated robot arm during VR assembly tasks}. \textbf{(2)} \textit{Practical design recommendations to enhance  multimodal interactions with NMUIs in assembly tasks}. \textbf{(3)} \textit{A dataset with annotated human interactions involving a simulated robot arm during VR assembly tasks which provides insights into how users interacted in both tasks}.

\section{Related Work}\label{section: related work}

In this section, we detail previous works relevant to our study: multimodal input systems and elicitation studies. We review recent works in these fields and identify gaps that our work aims to address.


\subsection{Multimodal Input Systems and Interfaces} \label{section: Multimodal Input Systems and Interfaces}
Multimodal Input Systems and Interfaces have been developed and used for decades, with fundamentals focused on how modalities interact with each other in more traditional interfaces like mouse and keyboard~\cite{oviatt2000perceptual, 10.1145/1027933.1027957, oviatt2003advances}. The paradigms of early, late, and hybrid fusion are detailed in these works in terms of how inputs are fused. More advanced interfaces include ones that integrate speech and gestures~\cite{cohen1997quickset, epps2004integration, johnston1997unification, oviatt2000designing}. These works use classification methods such as neural networks and typed feature structures to accept inputs in specific structures while allowing some degree of variability of input sequence. Findings from the use of these systems are based on spatio-temporal patterns such as inputs like speech and gesture being sequential rather than simultaneous, as well as statements being grounded in context by other relevant modalities. 

Recent natural interfaces have been designed for applications such as robot interaction~\cite{su2023recent} and VR~\cite{martin2022multimodality}, with emphasis also placed on task collaboration~\cite{ghamandi2023and, ghamandi2024unlocking}. This includes more recent work that has also been done to develop natural input models capable of learning the intricacies of these inputs. Our work extends prior research and differs from it by creating an environment to capture a wide array of sequences of multiple modalities, where users are untrained and complex multimodal instructions are accepted. Additionally, the sequences of modalities captured are context-aware and adaptive in order to account for previous interactions and mixed usage of modalities accordingly, with our work also exploring these system interactions in VR.

\subsection{Multimodal Datasets} \label{Multimodal Datasets}

Many datasets have been created during elicitation studies~\cite{10.1145/1518701.1518866, cohen2008high} by recording the natural interactions of users with the system for modalities such as speech, hand tracking, etc. with user perception in mind based on environment~\cite{hmaiti2024visual, hmaiti2023exploration}. These datasets serve the purpose of aiding trend analysis, visual data analysis via visualizations, as well as training machine learning models for inference based on data patterns. In the context of assembly tasks, many datasets have been created which capture the natural interactions between user and system. They include HARMONIC made for assistive human-robot collaboration~\cite{newman2022harmonic}, MMHRI for multimodal human-human-robot interactions~\cite{8003432}, and Hugging Robot for human-robot interactions for physical contact~\cite{bagewadi2019multimodal}. Other multimodal datasets that have been collected from natural human interaction studies in assembly include those leaning towards dyadic conversations~\cite{rozgic2010new, aubrey2013cardiff, marshall20154d, jayagopi2013vernissage}, as well as those for emotion and social instances~\cite{ringeval2013introducing, devillers2015multimodal, kesim2023ehri, inamura2021sigverse}, natural interaction~\cite{bilakhia2015mahnob, ben2017ue, shrestha2024natsgd, 10.1145/3568294.3580080, 10.1145/3610977.3636440, azagra2016multimodal, jiang2022vima, mahadevan2021grip}, among others~\cite{sanchez2011audio, 9223340}. Unlike these works, however, ours captures the natural interactions of untrained users in the form of voice, hand gestures, and head position and gaze while intuitively performing assembly tasks in VR. 

\section{Methodology}\label{Methodology}
The tasks employed are a LEGO Assembly Task, which we refer to as Task 1, and a PCB Assembly Task, which we refer to as Task 2. We employed these tasks as they varied in the type of collaboration used. In the following sections, we describe our Task Design, Apparatus, Study Design, Research Questions, Participant Demographics, and Procedure.

\begin{figure}[!h]
    \centering
    \includegraphics[width=\columnwidth]{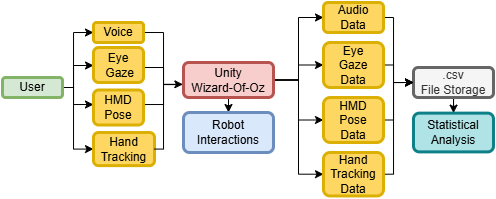}
    \caption{Flow diagram of data collection, processing, and storage for post-study statistical analysis.}
    \label{fig:collectionarchitecture}
\end{figure}

\subsection{Task Design} \label{Task Design}


The elicitation study was carried out in two separate instances; one with a LEGO assembly task, the other with a PCB assembly task. Both tasks were carried out with all objects and the robot arm laying on a virtual table, with the relevant reference assembly on the left of the participant and the area that they had to recreate the assembly and the robot arm being on their right. The summarized details for each task are shown in \cref{fig:TaskSpace}.

For each task, participants had to recreate one of two reference assemblies. They underwent no training in order to reveal intuitive and natural interactions with the system. The input modalities collected during each task were speech, gestures, eye gaze, and head pose.

Participants used varying combinations of the available input modalities to interact with the environment after being told the goal of the task. After this, the experimenter performing the Wizard-of-Oz interactions interpreted these instructions and controlled the robot in the background by listening to the user's voice as well as observing their hand movement and eye gaze and head position in the virtual environment (VE). Controls for the robot can be found in \cref{Apparatus}. 

In Task 1, participants collaborated with the robot by manipulating LEGO blocks and instructing the robot to pick and place pieces. The robot had its own set of LEGO separate from the user's. The LEGO task was chosen for collaboration as both the user and arm would be able to manipulate objects together in the real world. In this task setting, LEGO provide concrete spatial anchors that are easy to reference as locations for object placement when using speech.


For Task 2, participants had to instruct the robot on which components to assemble on the blank PCB board. Participants knew where to place components by looking at the precise spots on the reference PCB where reference components were (which were highlighted with visual outlines), after which they would know where to spatially indicate locations on the blank PCB. The PCB task was chosen as it reflects real-world applications where robot arms handle delicate electronic components with precision. Unlike LEGO, the PCB board lacks clear spatial anchors, making speech-based instructions more challenging.

For user error correction, if either the user or the robot placed an incorrect object based on the user’s request and the user wished to remove it, the experimenter would use the arm to select the object and press the 'Destroy' button to remove it from the scene. No errors occurred due to the Wizard-of-Oz operation, owing to the nature of the user interface (see ~\ref{Apparatus} for details).


It should be further noted that the selected tasks have simulated properties such as precise assembly and electrical compatibility, and are simplified versions of automated, mass-production real-world tasks. However, they were chosen to elicit the common behaviors associated with natural interaction with these systems in a realistic manner.


\subsection{Apparatus} \label{Apparatus}

A Meta Quest Pro VR HMD with a system developed in Unity Engine 2022.3.8f1 was used for the study. For the Wizard-of-Oz setup, the experimenter controlled the virtual robot arm using a keyboard for 3-axis movement (up and down arrow keys for x-axis, left and right arrow keys for y-axis, Q and E keys for z-axis) and a mouse for object manipulation (release, grip, move to, pick up, put down, destroy) and selecting discrete objects and locations for picking and placing, respectively. This allowed for precise object selection and placement without operational errors. A recording button was available to manage input capture. The data collection process is shown in \cref{fig:collectionarchitecture}. 

\subsection{Study Design} \label{Study Design}

\begin{figure}[!htb]
    \centering
    \includegraphics[width=\columnwidth]{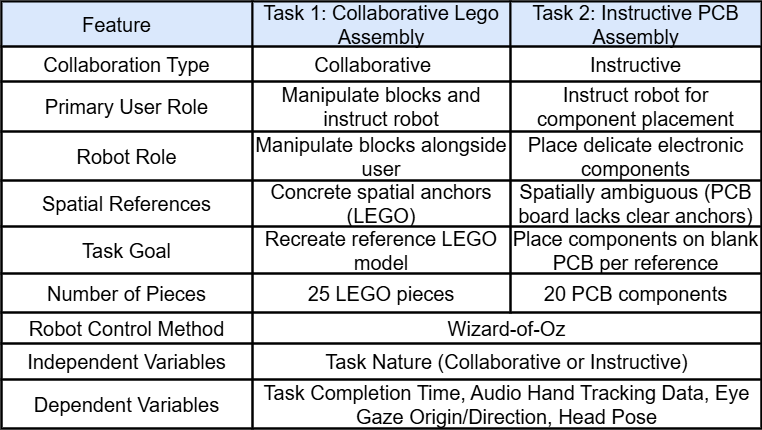}
    \caption{Task space for Task 1 (top) and Task 2 (bottom).}
    \label{fig:TaskSpace}
\end{figure}

To investigate individual behaviors and the nature of different types of assembly tasks, we conducted a single factor within-subjects study with two conditions: Collaborative Assembly (\textbf{Task 1}) and Instructive Assembly (\textbf{Task 2}). These two different assembly tasks were chosen in order to investigate what each varying task nature entails and to elicit how individuals complete tasks within these contexts. Counterbalancing was done between the two task types as well as the two reference assemblies; each participant was administered one of two task types first with one of two references administered for each task.

To avoid potential confounding variables (e.g. long-distance locomotion), participants had access to everything they needed for their part of the task in their immediate space. Participants also underwent training in order to learn what the task objectives were as well as the task mechanics of manual object manipulation in Task 1 in order to avoid any learning effects. However, participants did not receive training on how to interact with the robot or what it would respond to.

Objective dependent variables included task completion time, with other dependent data collected in the form of audio (for voice), hand tracking, eye gaze origin and direction, and head pose (see \cref{Dataset} for more details on collected data).



VR was used to study assembly, offering a controlled environment for data collection and mitigating safety concerns, complexities, and costs associated with physical robots \cite{laviola20173d, zhang2023research, husainy2023impact}. However, limitations include the inability to perform tasks in the physical world and the absence of haptic or tactile cues, which may restrict the transferability of findings related to fine motor control and error recovery to real-world scenarios. In contrast, higher-level behaviors, such as language use for spatial references and modality coordination, are transferable ~\cite{daling2024effects, carlson2015virtual}.

\subsection{Research Questions} \label{Research Questions}
Given the structure and parameters of our study, we propose the following research questions, which we base off of work from Bolt~\cite{bolt1980put} and Oviatt et al.~\cite{oviatt1997mulitmodal}, as well as from Shrestha et al.~\cite{shrestha2024natsgd}:

\begin{itemize}[left=5pt, itemsep=2pt, topsep=2pt]
    \item[] \textbf{Q1}: How do untrained users integrate voice, head gaze, hand tracking, and head pose when interacting with NMUIs in VR assembly tasks?
    \item[] \textbf{Q2}: How does the nature of assembly tasks influence the selection and effectiveness of input modalities in NMUIs?
    \item[] \textbf{Q3}: What strategies do untrained users adopt when using NMUIs to complete assembly tasks in VR?
\end{itemize}

\subsection{Participants} \label{Participants}
We recruited 34 participants for our study, all of which were students of the university and recruited through mass email/word of mouth. Participants were required to be 18 years of age or older, have normal or corrected-to-normal vision, and be able to hear, walk, extend both arms, use both hands, and speak and understand English. Participants with any visual, auditory, neurological, or physical disabilities were excluded. Our final participant pool comprised 34 individuals (21 male and 13 female). The ages of our participants ranged from 18 to 27 with a mean age of 21.23 and standard deviation of 2.44. The average user's experience with VR on a scale of 1 to 7 (with 7 being very experienced, 1 being no experienced) was 3.51, with a standard deviation of 1.60. 

\subsection{Procedure} \label{Procedure}



Upon arrival, participants were greeted by the experimenter, then filled out a consent form then demographics survey. After this, participants were given details about their first task. This was done by having the participant observe the VE on a computer screen while the experimenter wore the headset. The experimenter showed which objects in the task were manipulatable and the reference they had to recreate. After this, participants were given the VR headset to wear and how to adjust it, at which point the experimenter explained to them which input modalities the robot would interpret from them as they performed  the task. For Task 1, the experimenter allowed the participant to practice assembling LEGO before the task started. Participants were not aware that the robot arm was going to be controlled via Wizard-of-Oz. Each LEGO reference assembly had 25 pieces that had to be put together while each PCB assembly had 20 pieces that had to be placed on the PCB board. After completing both tasks, participants were thanked, compensated ten dollars for their time, and were allowed to leave. The study duration was approximately one hour.


\section{Dataset} \label{Dataset}


Our dataset, derived from our elicitation study, includes voice, gesture, eye gaze, and head pose data. Unity processed and stored 68 sessions (34 participants x 2 tasks) on a per-frame basis. Feature engineering (the formulation of new features from existing ones~\cite{dong2018feature}) was applied to extract new features for statistical analysis and application demonstrations. This dataset~\footnote{\url{https://osf.io/hbvxf/?view_only=b4347df6c0ea40a5a8d18ed1da0927e4}} is publicly available for research on natural multimodal interactions.

\subsection{Voice Data} \label{Dataset Voice Data}
Voice recordings were captured using the integrated microphone of the Meta Quest Pro headset.
The voice data was recorded in WAV format and were subsequently recordings processed using the speech-to-text model Whisper-Large Version 2~\cite{radford2023robust} to transcribe the audio into textual commands by timestamp, which were then saved in CSV format.

Additionally, the transcribed data was annotated to classify the types of commands given by the participants. The annotations were categorized into nine labels reflecting the command's intent and function: \textbf{(1)} \textbf{\textit{Pick and Place Explicit}}, \textbf{(2)} \textbf{\textit{Pick and Place Implicit}}, \textbf{(3)} \textbf{\textit{Pick Explicit}}, \textbf{(4)} \textbf{\textit{Pick Implicit}}, \textbf{(5)} \textbf{\textit{Place Explicit}}, \textbf{(6)} \textbf{\textit{Place Implicit}}, \textbf{(7)} \textbf{\textit{Arm Move Explicit}}, \textbf{(8)} \textbf{\textit{Arm Move Implicit}}, \textbf{(9)} \textbf{\textit{Unknown}}.
This annotation scheme was chosen based on Bolt's put-that-there~\cite{bolt1980put} in order to analyze the commands across tasks, with explicit commands corresponding to more descriptive commands (e.g. take a red block and place it on the orange block I put down just now) and implicit commands corresponding to more spatially vague commands (e.g. pick up this and put it here). The label distribution can be found in \cref{fig:Task Utterances}.


\subsection{Hand Tracking} \label{Dataset Hand Tracking}

Hand tracking data was captured with the use of the Meta XR SDK. Data capturing was managed by Unity, which provided raw tracking data in the form of position and rotation data in world space for each joint, which we refer to as the pose data. The data from both hands were collected, with the joints collected consisting of \textit{Hand\textunderscore Thumb3}, \textit{Hand\textunderscore Index1}, \textit{Hand\textunderscore Index3}, \textit{Hand\textunderscore Middle1}, \textit{Hand\textunderscore Pinky0}, \textit{Hand\textunderscore Pinky3}, \textit{Hand\textunderscore MiddleTip}, \textit{Hand\textunderscore RingTip}, and \textit{Body\textunderscore HandThumbTip}. These joints correspond to the thumb distal joint (Distal Phalanx of the Thumb), index finger metacarpal joint (Proximal Phalanx of the Index Finger), index finger distal joint (Distal Phalanx of the Index Finger), middle finger metacarpal joint (Proximal Phalanx of the Middle Finger), pinky finger base joint (Metacarpal of the Pinky Finger), pinky finger distal joint (Distal Phalanx of the Pinky Finger), tip of the middle finger (Distal Phalanx Tip of the Middle Finger), tip of the ring finger (Distal Phalanx Tip of the Ring Finger), and tip of the thumb (Distal Phalanx Tip of the Thumb), respectively.

From there, we used feature engineering to determine pointing gestures in both tasks. A pose was considered a point if the absolute difference of z-axis rotations of the \textit{Hand\textunderscore Index1} and \textit{Hand\textunderscore Index3} was below 10 degrees and if the difference of z-axis rotations of the \textit{Hand\textunderscore Middle1} and \textit{Hand\textunderscore MiddleTip} was above 30 degrees. Each pointing pose was counted for the user as a single point if the pose's duration lasted longer than 0.5 seconds. For example, if the pointing pose was detected for an uninterrupted sequence of more than 0.5 seconds, then one point was considered to be made.

\subsection{Eye Gaze Data}
\label{Eye Gaze Data}

The Meta Quest Pro headset coupled with the Meta XR SDK was used to capture eye gaze data from each participant. Prior to data collection, the eye tracking system was calibrated for each individual to ensure accuracy throughout the tasks. The data collection was conducted via Unity, which recorded the eye movements as raw data during both assembly tasks. This was captured in the form of both position and orientation data in world space for the origin and direction for each ray for each eye. 

From here, we used feature engineering to determine whether a participant is looking left or right in the environment (gaze direction) using head pose data coupled with Eye Gaze data. Gaze direction was determined by using the origin and direction of the eye gaze to know whether the person was looking towards the side with the reference or the assembly, where the head pose was used to determine where the person was in world space.

\subsection{Head Pose Data}
\label{Dataset HMD Data}



Head Pose data was collected by logging the position and orientation of the HMD in world space. This data was collected in order to determine where the user was looking in the environment. In addition to this, we used feature engineering to determine whether the participant's head faced left (the general area with the reference assembly) or right (the general area with the assembly area) in relevance to their starting point in both the tasks, which was between the assembly reference and the actual workspace with the robot arm.

As described in \cref{Eye Gaze Data}, head pose data was coupled with Eye Gaze data to determine gaze direction. In addition, feature engineering was used to determine positional and rotational velocity. Positional velocity was obtained by calculating the euclidean difference between the position of consecutive time points and dividing by the difference in time; rotational velocity was obtained in the same way but quaternions between consecutive timestamps were used instead.


\section{Results} \label{Results}

In the following section, we report the detailed analysis of the data recorded for further understanding the nuances of our dataset collected for natural interaction across different modalities. Also, we have grouped the results based on statistical tests for Wilcoxon (See \cref{tab: wilcoxon}), Coefficient of Multimodal Correlation (See \cref{tab: Coefficient}), Paired Samples T-test (See \cref{tab: t-test}), and Unimodal Correlation (See \cref{tab: Correlation}). Before conducting specific statistical tests, the normality of the data for each variable was assessed using the Shapiro-Wilk test. If the data met the assumption of normality, paired samples t-tests or Pearson correlation tests were applied. For data that violated the assumption of normality, non-parametric alternatives such as the Wilcoxon Signed-Rank test or Spearman correlation were used. This procedure was consistently applied across all analyses for voice, hand tracking, eye gaze, and multimodal interactions. For paired samples t-tests and Wilcoxon Signed-Rank tests, data from Task 1 and Task 2 were compared to determine the effect of Task Nature (collaborative, concrete spatial references vs. instructive, ambiguous spatial references) on modality usage and tendencies. These tests were done to investigate each of our research questions, with the intent being to learn what patterns come about in these tasks, what trends exist that affect task completion, and identify what strategies users take up to perform tasks.

\begin{table}[ht]
\caption{Table showing Wilcoxon Signed-Rank test results determining the differences in specific modality usages and behaviors for Voice, Hand, and Eye Gaze modalities due to task nature (\textit{* = $p < .05$; ** = $p < .01$; *** = $p < .001$})}
\label{tab: wilcoxon}
\resizebox{\columnwidth}{!}{%
\begin{tabular}{|l|l|l|l|l|l|l|l|l|}
\hline
\textbf{\textit{Modality}} & \textbf{\textit{Name}} & \textbf{\textit{Z}} & \textbf{\textit{p}} & \textbf{\textit{Sig}} & \textbf{\textit{$M_1$}} & \textbf{\textit{$SD_1$}} & \textbf{\textit{$M_2$}} & \textbf{\textit{$SD_2$}} \\ \hline
Eye Gaze & Avg. Time Looking Right     & 242.00 & 0.35    & No  & 6.56   & 2.01    & 7.56   & 2.97    \\ \hline
Voice    & Phase 1 Labels (Task 1)     & -2.30  & 0.02    & *   & 1.21   & 1.43    & 0.56   & 0.86    \\ \hline
Voice    & Phase 2 Labels (Task 1)     & -2.29  & 0.02    & *   & 4.18   & 2.89    & 2.26   & 2.30    \\ \hline
Voice    & Phase 3 Labels (Task 1)     & 0.17   & -1.37   & No  & 1.03   & 1.03    & 0.79   & 1.34    \\ \hline
Voice    & Phase 1 Labels (Task 1)     & -3.33  & \textless{}0.001  & *** & 0.50   & 1.23    & 2.23   & 1.28    \\ \hline
Voice    & Phase 2 Labels (Task 1)     & -4.25  & \textless{}0.001  & *** & 1.06   & 2.59    & 6.88   & 2.50    \\ \hline
Voice    & Phase 3 Labels (Task 1)     & -3.97  & \textless{}0.001  & *** & 0.38   & 0.78    & 2.15   & 1.46    \\ \hline
Hand     & Average Point Time          & 37.00  & \textless{}0.001  & *** & 1.28   & 0.61    & 3.48   & 2.14    \\ \hline
Hand     & Q2-Q1 Contrast (Task 1)     & -1.14  & 0.26    & No  & 0.24   & 0.15    & 0.30   & 0.33    \\ \hline
Hand     & Q3-Q1 Contrast (Task 1)     & -1.75  & 0.08    & No  & 0.24   & 0.15    & 0.34   & 0.30    \\ \hline
Hand     & Q4-Q1 Contrast (Task 1)     & -2.49  & 0.01    & *   & 0.24   & 0.15    & 0.39   & 0.37    \\ \hline
Hand     & Q3-Q2 Contrast (Task 1)     & -0.62  & 0.53    & No  & 0.30   & 0.33    & 0.34   & 0.30    \\ \hline
Hand     & Q4-Q2 Contrast (Task 1)     & -1.75  & 0.08    & No  & 0.30   & 0.33    & 0.39   & 0.37    \\ \hline
Hand     & Q4-Q3 Contrast (Task 1)     & -0.88  & 0.38    & No  & 0.34   & 0.30    & 0.39   & 0.37    \\ \hline
Hand     & Q2-Q1 Contrast (Task 2)     & -0.98  & 0.33    & No  & 1.23   & 1.00    & 1.63   & 1.93    \\ \hline
Hand     & Q3-Q1 Contrast (Task 2)     & -1.09  & 0.28    & No  & 1.23   & 1.00    & 1.33   & 1.14    \\ \hline
Hand     & Q4-Q1 Contrast (Task 2)     & -1.50  & 0.14    & No  & 1.23   & 1.00    & 1.37   & 1.18    \\ \hline
Hand     & Q3-Q2 Contrast (Task 2)     & -0.25  & 0.80    & No  & 1.63   & 1.93    & 1.33   & 1.14    \\ \hline
Hand     & Q4-Q2 Contrast (Task 2)     & -0.42  & 0.68    & No  & 1.63   & 1.93    & 1.37   & 1.18    \\ \hline
Hand     & Q4-Q3 Contrast (Task 2)     & -0.04  & 0.97    & No  & 1.33   & 1.14    & 1.37   & 1.18    \\ \hline
\end{tabular}
}
\end{table}

\begin{table}[]
\caption{Table showing coefficient of Multimodal Correlation Analyses determining the effect of Utterance Frequency, Average Pointing Time, And Gaze Stationary Time Ratio on Task Completion Time (\textit{* = $p < .05$; ** = $p < .01$; *** = $p < .001$})}
\label{tab: Coefficient}
\resizebox{\columnwidth}{!}{%
\begin{tabular}{|l|l|l|l|l|l|l|}
\hline
\textbf{\textit{Modality}} & \textbf{\textit{Name}} & \textbf{\textit{F}} & \textbf{\textit{p}} & \textbf{\textit{Sig}} & \textbf{\textit{R}} & \textbf{\textit{$R^2$}} \\ \hline
Multimodal & Multimodal Correlation (Task 1) & $F_{3, 30} = 5.438$  & $< .01$          & **     & 0.535 & 0.287                \\ \hline
Multimodal & Multimodal Correlation (Task 2) & $F_{3, 30}$ = 27.62 & $< .001$         & ***    & 0.841 & 0.708                \\ \hline
\end{tabular}
}
\end{table}

\begin{table}[ht]
\caption{Table showing paired samples T-Tests determining the differences in specific Modality Usages and Behaviors for Voice and Eye Gaze modalities due To Task Nature as well as Multimodal Input Usages (\textit{* = $p < .05$; ** = $p < .01$; *** = $p < .001$})}
\label{tab: t-test}
\resizebox{\columnwidth}{!}{%
\begin{tabular}{|l|l|l|l|l|l|l|l|l|}
\hline
\textbf{\textit{Modality}} & \textbf{\textit{Name}} & \textbf{\textit{t}} & \textbf{\textit{p}} & \textbf{\textit{Sig}} & \textbf{\textit{$M_1$}} & \textbf{\textit{$SD_1$}} & \textbf{\textit{$M_2$}} & \textbf{\textit{$SD_2$}} \\ \hline
Voice         & Implicit Utterances Ratio                       & -8.14 & $< .001$         & ***  & 0.37   & 0.31  & 0.82   & 0.24   \\ \hline
Voice         & Explicit Utterances Ratio                        & 8.14  & $< .001$         & ***  & 0.63   & 0.31  & 0.18   & 0.24   \\ \hline
Eye Gaze      & Average Time Looking Left                   & 3.2   & $< .01$          & **   & 2.60   & 0.59  & 2.20   & 0.47   \\ \hline
Multimodal    & Unimodal Modality Usage                     & 2.78  & $< .01$          & **   & 0.24   & 0.17  & 0.15   & 0.13   \\ \hline
Multimodal    & Bimodal Modality Usage                      & 1.91  & 0.064            & No   & 0.53   & 0.18  & 0.43   & 0.19   \\ \hline
Multimodal & Trimodal Modality Usage        & -5.07 & $< .001$         & ***    & 0.15  & 0.21  & 0.42  & 0.24   \\ \hline
\end{tabular}
}
\end{table}

\begin{table}[ht]
\caption{Table showing correlation analyses determining the relationship between Utterance Frequency, Average Pointing Time, Stationary Time Ratio and Task Completion Time (\textit{* = $p < .05$; ** = $p < .01$; *** = $p < .001$})}
\label{tab: Correlation}
\resizebox{\columnwidth}{!}{%
\begin{tabular}{|l|l|l|l|l|l|l|l|l|}
\hline
\textbf{\textit{Modality}} & \textbf{\textit{Name}} & \textbf{\textit{r}} & \textbf{\textit{p}} & \textbf{\textit{Sig}} & \textbf{\textit{$M_1$}} & \textbf{\textit{$SD_1$}} & \textbf{\textit{$M_2$}} & \textbf{\textit{$SD_2$}} \\ \hline

Voice         & Utterance Frequency (Task 1)   & 0.39   & $< .05$    & *    & 11.94  & 5.07   & 382.84  & 148.7   \\ \hline
Voice         & Utterance Frequency (Task 2)   & 0.85   & $< .05$    & *    & 17.32  & 5.67   & 451.4   & 133.77  \\ \hline
Hand & Avg. Pointing Time (Task 1)    & 0.18   & 0.32       & No   & 1.93   & 1.00   & 382.84  & 148.7   \\ \hline
Hand & Avg. Pointing Time (Task 2)    & 0.15   & 0.41        & No   & 4.24   & 2.42   & 451.4   & 133.77  \\ \hline
Eye Gaze      & Stationary Time Ratio (Task 1) & 0.37   & $< .05$    & *    & 0.29   & 0.12   & 382.84  & 148.7   \\ \hline
Eye Gaze      & Stationary Time Ratio (Task 2) & -0.025 & 0.88       & No   & 0.39   & 0.14   & 451.4   & 133.77  \\ \hline
\end{tabular}
}
\end{table}

\subsection{Voice Data} \label{Results Voice Data}

For our analysis of voice, we first quantified the number of utterances for each participant based on the labels we chose. As stated before, explicit utterances refer to those that are descriptive and clear, while implicit utterances refer to those that are spatially vague and benefit from additional spatial cues. Further details can be found in \cref{Dataset Voice Data}. Excluding the unknown utterances, \textbf{\textit{Pick and Place Explicit}} was found to be the predominant annotation in Task 1, while \textbf{\textit{Pick and Place Implicit}} was found to be the predominant annotation in Task 2.

For our statistical analyses, we conducted multiple analyses across tasks involving the assigned labels.

\subsubsection{Explicit/Implicit Utterances Ratio} \label{section: Explicit/Implicit Utterances Ratio}
\begin{figure}[!h]
    \centering
    \includegraphics[width=\columnwidth]{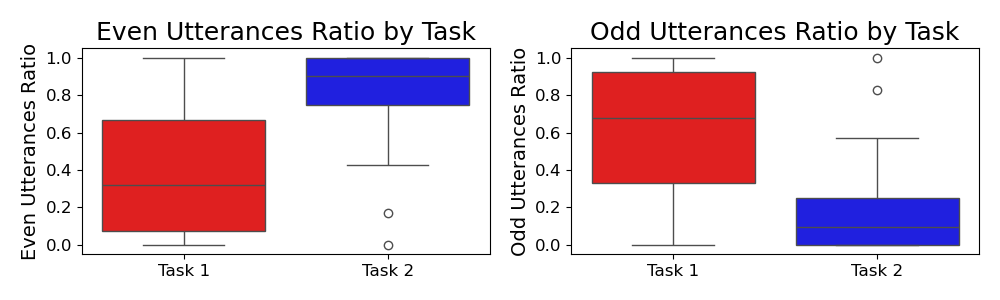}
    \caption{Ratios of Explicit and Implicit Utterances. More Explicit Utterances were used in Task 1, while more Implicit Utterances were used in Task 2.}
    \label{fig:evenoddutterances}
\end{figure}

Explicit utterances refer to utterances that fall under the labels of 1, 3, 5, or 7. Meanwhile, implicit utterances refer to utterances that fall under the labels of 2, 4, 6, or 8. Ratios were calculated for each participant by summing the number of utterances falling under each type of label (implicit or explicit) and dividing by the total number of utterances excepting those that were \textbf{\textit{9 - Unknown}}.


A paired samples t-test revealed that participants used significantly more implicit utterances in Task 2 ($M = 0.82, SD = 0.24$) compared to Task 1 ($M = 0.37, SD = 0.31$), $t_{33} = -8.14, p < 0.001$. Conversely, explicit utterances were significantly more frequent in Task 1 ($M = 0.63, SD = 0.31$) than in Task 2 ($M = 0.18, SD = 0.24$), $t_{33} = 8.14, p < 0.001$. A summary of these analyses is provided in \cref{tab: t-test}, and the distribution of utterance ratios is shown in \cref{fig:evenoddutterances}.


\begin{figure*}[!tbp]
    \centering
    \includegraphics[width=.9\textwidth]{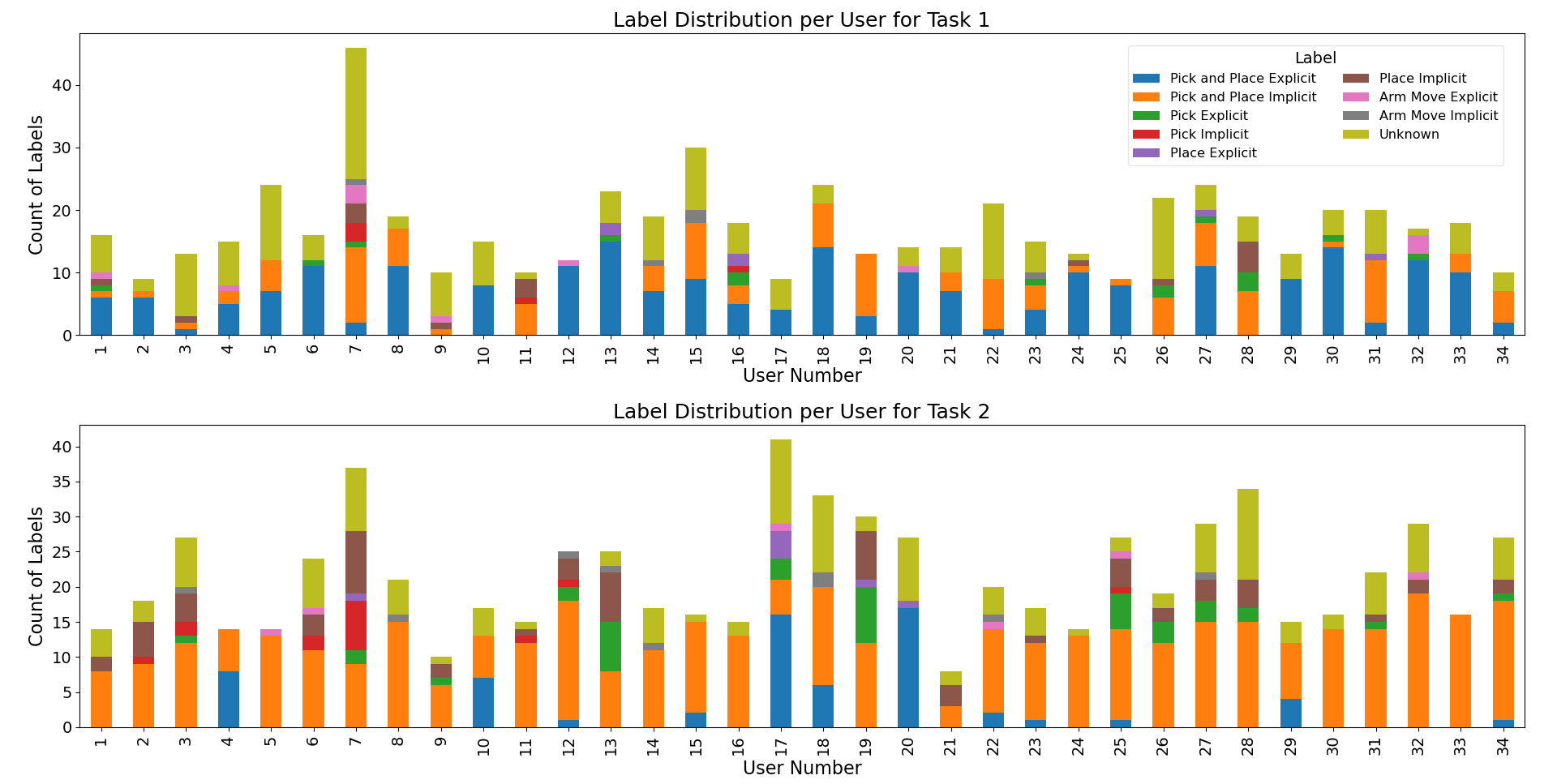}
    \caption{Stacked bar graphs show the distribution of utterance labels across the two tasks for each participant. \textit{\textbf{Pick and Place Explicit}} was used most frequently in Task 1, while \textit{\textbf{Pick and Place Implicit}} was used most frequently in Task 2.}
    \label{fig:Task Utterances}
\end{figure*}

\subsubsection{Utterance Frequency and Task Completion Time Correlation} \label{section: Utterance Frequency Correlation}
We conducted a linear correlation analysis in order to determine the effect of utterance frequency on task completion time for Task 1 and Task 2. 




For Task 1, a Spearman correlation test revealed a moderate positive correlation between utterance frequency ($M = 11.94, SD = 5.07$) and Task Completion Time ($M = 382.84, SD = 148.70$), $r_{32} = 0.39, p = 0.022$. For Task 2, a Pearson correlation test indicated a strong positive correlation between utterance frequency ($M = 17.32, SD = 5.67$) and Task Completion Time ($M = 451.40, SD = 133.77$), $r_{32} = 0.85, p = 0.022$. A summary of these analyses is provided in \cref{tab: Correlation}.

\subsubsection{Utterance Usage Per Task Phase} \label{section: Utterance Usage Per Task Phase}
\begin{figure}[!h]
    \centering
    \includegraphics[width=\columnwidth]{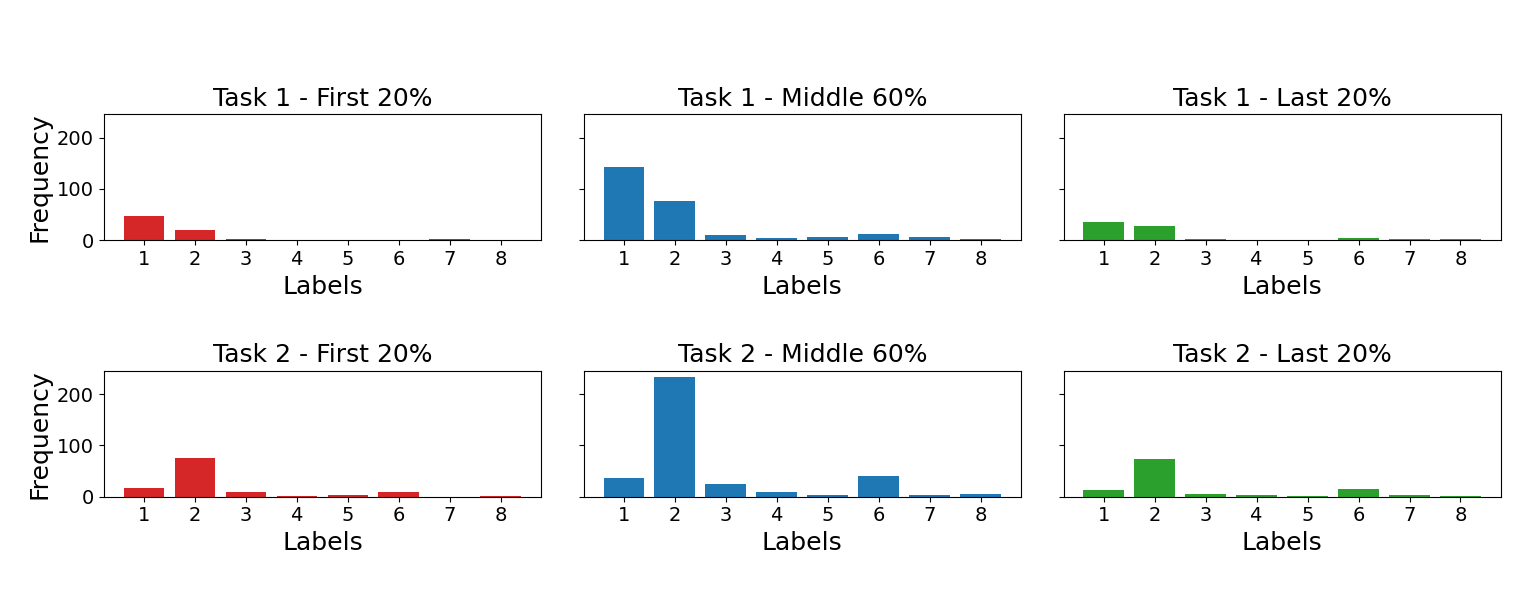}
    \caption{Distributions of Explicit and Implicit Utterances across Task Phases.}
    \label{fig:utterancesinphases}
\end{figure}

The frequency of utterances for different phases of each task were analyzed to determine how utterance usage changed over time. The utterances were grouped by different task completion percentages which defined phases 1, 2 and 3 respectively: 0\% to 20\%, 20\% to 80\%, and 80\% to 100\%.

We used a Wilcoxon Signed-Rank Test to test for significant differences in the usage between labels 1 and 2 for each phase as they were the most used utterances. For Task 1, the test indicated significant differences for phase 1 ($M_1 = 1.21, SD_1 = 1.43, M_2 = 0.56, SD_2 = 0.86, t_{33} = -2.30, p  = 0.021$) and phase 2 ($M_1 = 4.18, SD_1 = 2.89, M_2 = 2.26, SD_2 = 2.30, t_{33} = -2.29, p  = 0.022$), but not for phase 3 ($M_1 = 1.03, SD_1 = 1.03, M_2 = 0.79, SD_2 = 1.34, t_{33} = -1.37, p  = 0.17$). For Task 2, the test indicated significant differences for phase 1 ($M_1 = 0.5, SD_1 = 1.23, M_2 = 2.23, SD_2 = 1.28, t_{33} = -3.33, p < 0.001$), phase 2 ($M_1 = 1.06, SD_1 = 2.59, M_2 = 6.88, SD_2 = 2.50, t_{33} = -4.25, p < 0.001$), and phase 3 ($M_1 = 0.38, SD_1 = 0.78, M_2 = 2.15, SD_2 = 1.46, t_{33} = -3.97, p < 0.001$). 

A summary of these analyses can be seen in \cref{tab: wilcoxon}, while the distribution of utterance usage per phase can be found in \cref{fig:utterancesinphases}.


\subsection{Hand Tracking Data} \label{Results Hand Tracking Data}
Hand tracking data was analyzed to determine if task nature had an effect on how users interacted with the system in terms of performing specific actions and how. 


\begin{figure}[!h]
    \centering
    \includegraphics[width=.8\columnwidth]{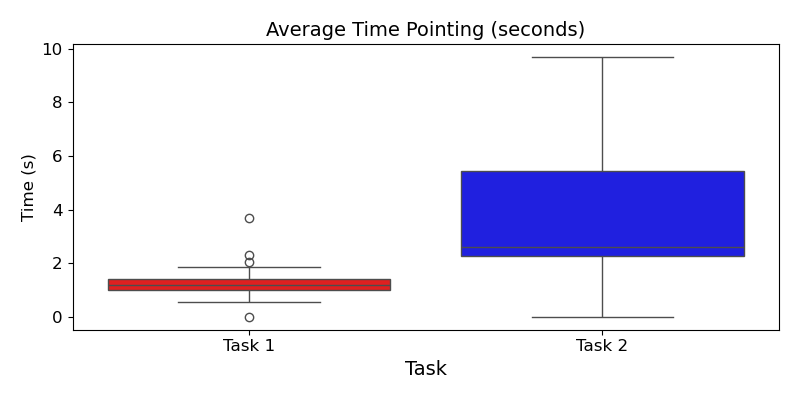}
    \caption{Graphs showing Average Pointing Times per Task. Participants pointed significantly longer on average in Task 2.}
    \label{fig:pointplots}
\end{figure}



\subsubsection{Average Point Time} \label{section: Average Pointing Time}
The average point time per task was analyzed to determine if task nature had an effect on how long people pointed to locations in the environment. Average point time was calculated by adding up the times of all pointing poses that lasted over 0.5 seconds and then dividing by the number of those points.


A Wilcoxon Signed-Rank Test was conducted to examine whether task type affected average pointing time. The test revealed a significant difference, with participants pointing longer in Task 2 ($M = 3.48, SD = 2.14$) than in Task 1 ($M = 1.28, SD = 0.61, t_{33} = 37.0, p < 0.001$).  A summary of this analysis is provided in \cref{tab: wilcoxon}, and the distribution of Average Point Times is shown in \cref{fig:pointplots}.

\subsubsection{Average Point Time and Task Completion Time Correlation}
We conducted a linear correlation analysis in order to determine the effect of average time spent pointing on task completion time for Task 1 and Task 2. 



Spearman correlation tests found no significant relationship between Average Pointing Time and Task Completion Time in either Task 1 ($M = 1.28, SD = 0.61$; $M = 382.84, SD = 148.70$; $r_{32} = 0.077, p = 0.66$) or Task 2 ($M = 3.48, SD = 2.14$; $M = 451.40, SD = 133.77$; $r_{32} = 0.14, p = 0.43$).

\subsubsection{Average Pointing Time Per Task Phase}
\begin{figure}[!h]
    \centering
    \includegraphics[width=\columnwidth]{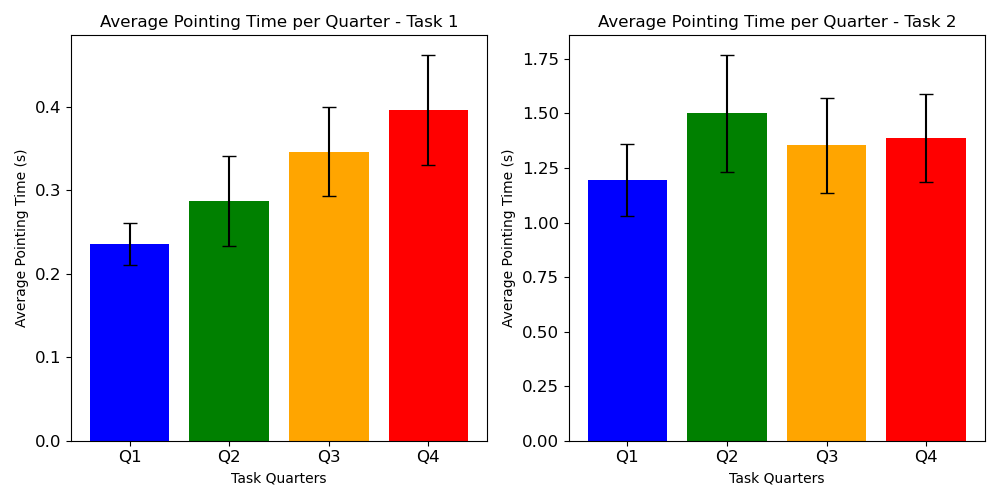}
    \caption{Average Pointing Times per Task Quarter for each Task. Average Pointing Time had an upward trend in Task 1, while there was no clear trend in Task 2. Error bars show 1 SEM (standard error of the mean).}
    \label{fig:handtrackinginphases}
\end{figure}

The average pointing time for different phases of each task were analyzed to determine the change in how much time users spent performing a point. Instead of using the threshold of a point lasting for at least 0.5 seconds, no threshold was used to determine if there were significant evolutions in how participants made points.


We used Wilcoxon Signed-Rank Tests to test for significant differences in average pointing time across quarters for each task to determine if the nature of the task affected the evolution of pointing averages. Of all tests conducted, significant differences were only found between Q1 ($M = 0.24, SD = 0.15$) and Q4 ($M = 0.39, SD = 0.37, t_{33} = -2.49, p = 0.013$), for Task 1.  A full report of these tests can be found in \cref{tab: wilcoxon}, while the distribution of average points per task phase for each task is shown in \cref{fig:handtrackinginphases}.

\subsection{Eye Gaze and Head Pose Data} \label{section: Results Eye Gaze and HMD Data}
Eye gaze and head pose data was analyzed to determine if task nature had an effect on how and when users looked at the reference assembly (left) and task workspace (right). This distinction was made by the starting position of the user in the environment, with the reference being on the left and the workspace being on the right.

\begin{figure}[!h]
    \centering
    \includegraphics[width=\columnwidth]{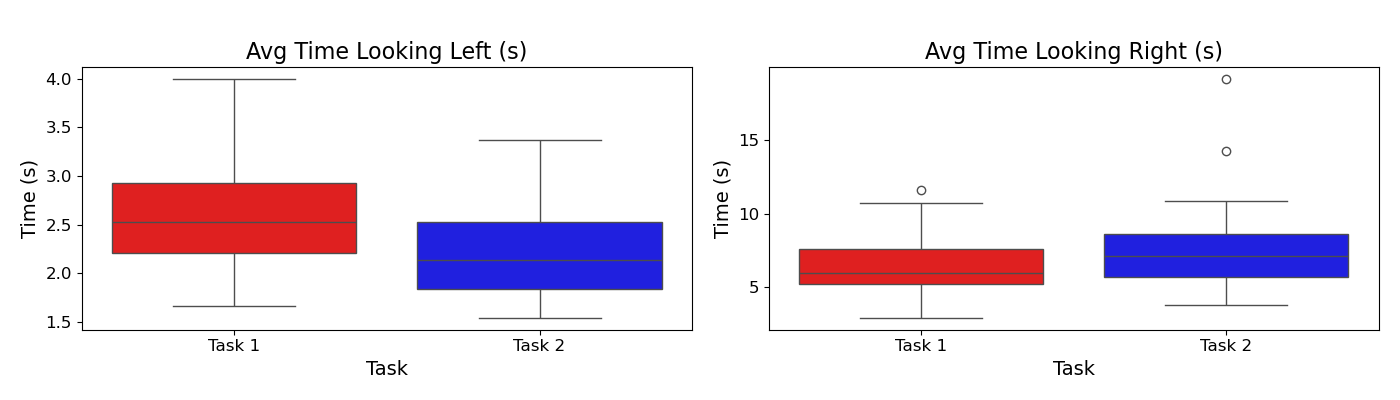}
    \caption{Average Look Durations for Task 1 and Task 2. Participants looked significantly longer to the left in Task 1. Meanwhile, there was no significant difference in how long participants looked to the right.}
    \label{fig:gazeplots}
\end{figure}

\subsubsection{Average Time Looking Left}

A paired samples t-test revealed that participants spent significantly more time looking left in Task 1 ($M = 2.60, SD = 0.59$) compared to Task 2 ($M = 2.20, SD = 0.47$, $t_{33} = 3.20, p = 0.003$). A summary of this analysis can be found in \cref{tab: t-test}.



\subsubsection{Average Time Looking Right}

A Wilcoxon Signed-Rank test revealed no significant difference in time spent looking right between Task 1 ($M = 6.56, SD = 2.01$) and Task 2 ($M = 7.56, SD = 2.97$, $t_{33} = 242.0, p = 0.35$). A summary of this analysis can be found in \cref{tab: wilcoxon}.

\subsubsection{Stationary Time Ratio and Task Completion Time Correlation}
We conducted a linear correlation analysis in order to determine the effect of stationary gaze ratios on task completion time for Task 1 and Task 2. 



For Task 1, a Spearman correlation revealed a significant positive relationship between stationary time ratio ($M = 0.29, SD = 0.12$) and Task Completion Time ($M = 382.84, SD = 148.70$, $r_{32} = 0.37, p = 0.031$). For Task 2, a Pearson correlation showed no significant relationship between stationary time ratio ($M = 0.39, SD = 0.14$) and Task Completion Time ($M = 451.40, SD = 133.77$, $r_{32} = -0.025, p = 0.88$). A summary of these analyses can be found in \cref{tab: Correlation}.

\subsection{Multimodal Analysis}


We analyzed how modalities (voice, gestures, and eye/head gaze) were used concurrently and how this interplay influenced task completion times, depending on task nature.

\subsubsection{Sequential Pattern Mining}
\begin{figure*}[!tbp]
    \centering
    \includegraphics[width=\textwidth]{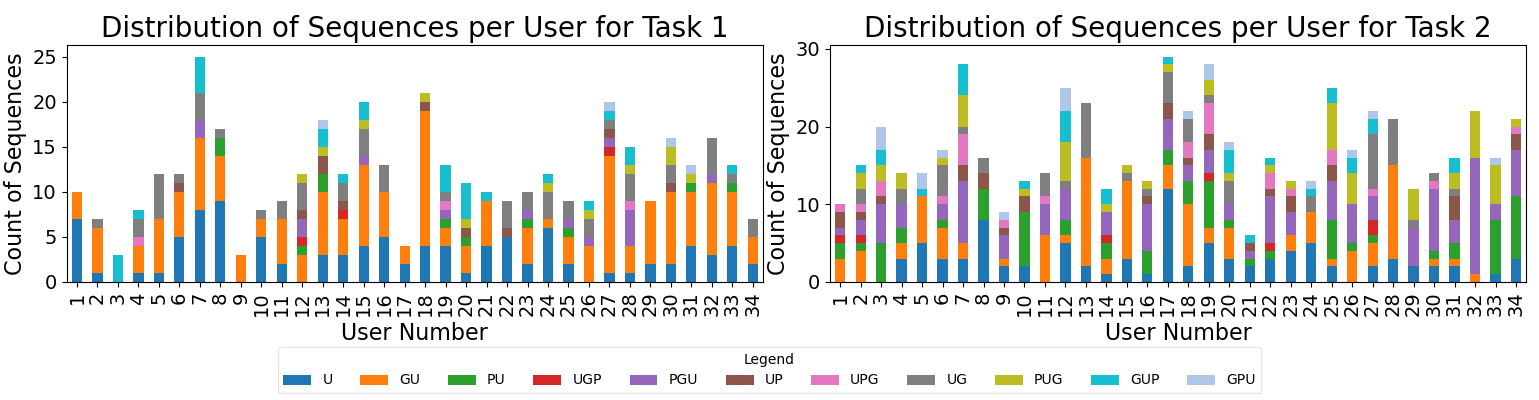}
    \caption{Stacked Bar Graph showing the Task Sequence Distribution across Task 1 (left) and Task 2 (right) for all participants (G -- Gaze, P -- Pointing Gesture, U -- Utterance).}
    \label{fig:Task Sequences}
\end{figure*}

Sequential Pattern Mining was performed to determine the sequences in which participants delivered multimodal inputs to the system to execute instructions. These patterns were centered around voice inputs that fell under one of the eight labels pre-defined. A sequence was recorded when an utterance was detected, after which stationary gaze and/or pointing gestures occurring within a five-second range of the start time of the utterance would comprise the sequence. The distribution of sequences for Task 1 and Task 2 can be found in \cref{fig:Task Sequences}.

\subsubsection{Unimodal, Bimodal, and Trimodal Comparisons}\label{section: Modal Analysis}

\begin{figure}[!tbp]
    \includegraphics[width=\columnwidth]{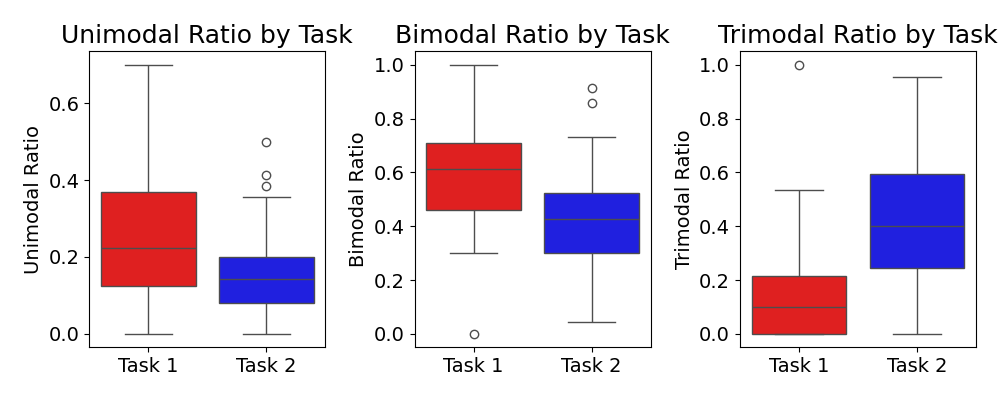}
    \caption{Unimodal, Bimodal and Trimodal Command usages per Task. Unimodal commands were used significantly more in Task 1, Bimodal command usage did not differ between tasks, and Trimodal commands were used significantly more in Task 2.}
    \label{fig:unibitriplots}
\end{figure}





Analyses were performed on unimodal, bimodal, and trimodal sequence frequencies to determine the effect of task nature on their uses. A paired samples t-test revealed that participants used significantly more unimodal sequences in Task 1 ($M = 0.25, SD = 0.19$) than in Task 2 ($M = 0.16, SD = 0.13$, $t_{33} = 2.89, p = 0.006$), and significantly more bimodal sequences in Task 1 ($M = 0.59, SD = 0.20$) than in Task 2 ($M = 0.42, SD = 0.21$, $t_{33} = 3.05, p = 0.004$). A Wilcoxon Signed-Rank test showed participants used significantly more trimodal sequences in Task 2 ($M = 0.42, SD = 0.24$) than in Task 1 ($M = 0.15, SD = 0.21$, $t_{33} = -5.06, p < 0.001$). A summary of these analyses can be found in \cref{tab: t-test}, and the distributions are shown in \cref{fig:unibitriplots}.

\subsubsection{Multimodal Correlation} \label{section: Multimodal Correlation}
A multimodal correlation analysis was conducted using a Multiple Correlation Coefficient test to determine the effect of utterance frequency, average pointing time and gaze stationary time ratio (the ratio of time each user's gaze was stationary compared to moving) on task completion time for both Task 1 and Task 2. A summary of the following tests can be found in \cref{tab: Coefficient}. For Task 1, the test indicated that there was a significant positive correlation ($F_{3, 30} = 5.296, p < 0.001, R = 0.530, R^2 = 0.281$). For Task 2, there was also a significant positive correlation ($F_{3, 30} = 27.63, p < 0.001, R = 0.841, R^2 = 0.708$).

\subsubsection{Sequence Correlation} \label{section: Sequence Correlation}
A multimodal correlation analysis was conducted using a Multiple Correlation Coefficient test to determine the effect of all unique sequence frequencies on task completion time for both Task 1 and Task 2. For Task 1, the test indicated that there was no significant correlation ($F_{11, 22} = 1.971, p = 0.085, R = 0.494, R^2 = 0.244$). For Task 2, there was a significant positive correlation ($F_{11, 22} = 7.874, p < 0.001, R = 0.834, R^2 = 0.696$).

\section{Discussion}\label{Discussion}
\begin{figure*}[!htb]
    \centering
    \includegraphics[width=\textwidth]{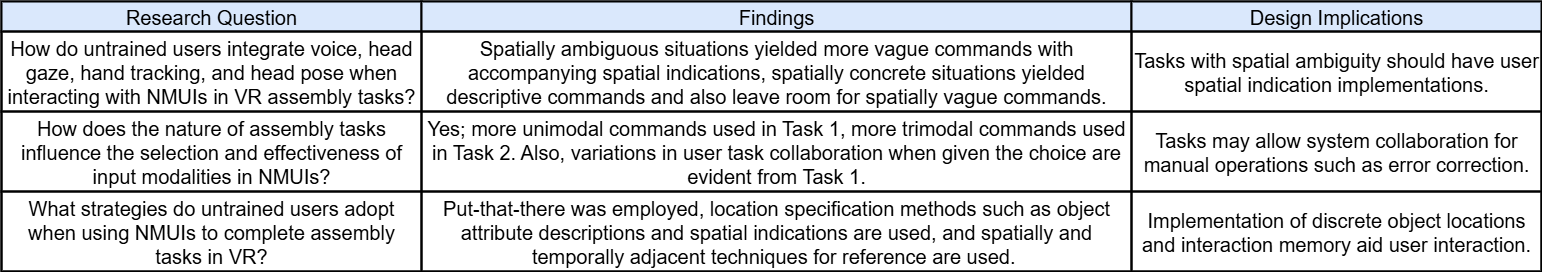}
    \caption{Task space for Task 1 (top) and Task 2 (bottom).}
    \label{fig:TaskSpace}
\end{figure*}




\subsection{Spatio-temporal Patterns} 

Our first research question asks how the various modalities used are integrated spatially and temporally for delivery to NMUIs during assembly tasks in VR based on the nature of the task by untrained users. 

Our findings show that in Task 1, users mainly combined descriptive utterances with quick nonverbal spatial indications and temporal references to recently manipulated objects (e.g. "blue block on top of red block", with the red block having been recently placed). Meanwhile, in Task 2, ambiguous utterances followed by nonverbal spatial indications (prolonged over time), were mainly employed (e.g. "take a resistor and put it here", with the word "here" being accompanied by a prolonged point). These trends are indicated in analyses performed on modality usage (see ~\cref{section: Modal Analysis}, \cref{fig:unibitriplots}), which show unimodal commands used significantly more in Task 1 and trimodal commands used significantly more in Task 2, but bimodal commands not significantly differing in usage between the tasks. Furthermore, tests on the average pointing time show significantly longer point times in Task 2 than Task 1 (see \cref{section: Average Pointing Time}, \cref{fig:pointplots}), followed by implicit utterances being used more in Task 2 than Task 1. These trends can be largely attributed to Task 1 having concrete spatial anchors which required little spatial indication, while Task 2 had ambiguous ones, even if some objects were adjacent, which prompted participants to prolong their spatial indications to ensure the robot received enough indications of their intent. These patterns align with work from Oviatt et al.~\cite{oviatt1997integration, 10.1145/319382.319398}, which explains that spatial context is elaborated with other modalities such as pen and gestures, while voice provides description for an instruction.

Other trends found include participants consistently using explicit commands while increasing their usage of implicit commands over time in Task 1 (e.g. \textbf{Pick and Place Implicit} usage increased for participants 8 and 15 over time). However, in Task 2, participants consistently use implicit commands from start to finish (e.g. participants 16 and 22 had no major trend in pointing usage with \textbf{Pick and Place Implicit} used throughout the task). These trends are apparent in analyses performed on utterance usage over task duration (see \cref{section: Utterance Usage Per Task Phase}, \cref{fig:utterancesinphases}) and average pointing times over task duration (see \cref{section: Average Pointing Time}, \cref{fig:pointplots}). They indicate that Task 1 facilitates the use of both implicit and explicit utterances, but Task 2 mainly facilitates implicit instructions only. This contrasts prior work from Oviatt et al.~\cite{oviatt1997integration, 10.1145/319382.319398} in that modality choice remains consistent throughout tasks, but aligns in that users adopt specific modality choices based on task type despite users changing preference over time in Task 1.

These spatio-temporal differences bear inter dependencies, with nonverbal cues dominating in situations in precision and usage due to ease of use and efficiency. Voice otherwise provides context and nuanced elaborations for delivering complex instructions.

\subsection{Task Nature Trends}

Our second research question explores how the nature of the assembly task influences the choice and effectiveness of modalities in NMUIs. First, more explicit commands were used in Task 1 and more implicit commands were used in Task 2. In addition, Task 2 completion time was strongly attributed to modality and utterance usage, but not Task 1 completion time.

For command usage in Task 1, descriptive utterances with brief spatial explanations were employed due to concrete spatial anchors being present, while vague utterances reinforced by prolonged spatial commands were used due to a lack of spatial anchors. This is evident in analyses on utterance distribution (see \cref{fig:Task Utterances}, \cref{section: Explicit/Implicit Utterances Ratio}), pointing time (significantly longer pointing times on average in Task 2 than Task 1 which shows longer spatial anchoring, see \cref{section: Average Pointing Time}, \cref{fig:pointplots}), and modality usage (more unimodal commands in Task 1, more trimodal commands in Task 2, see \cref{section: Modal Analysis}, \cref{fig:unibitriplots}).

Furthermore, Task 2 completion time being strongly, positively correlated to the frequencies of instructions (utterances, see \cref{section: Utterance Frequency Correlation}), delivered modality combinations, (sequences, see \cref{section: Sequence Correlation}) and the usage of multiple modalities (multimodal correlation, see \cref{section: Multimodal Correlation}) is due to users relaying multiple object placements in a single instruction. As Task 1 had comparatively weaker correlations, with a relatively low variance in multimodal correlation performed compared to Task 2 (see \cref{section: Multimodal Correlation}), this implies that frequencies of manual manipulation and robot instruction varied due to individual user preference of object manipulation method, further showing that they contribute to task efficiency, leading to design considerations in terms of allowing users to deliver instructions via preferred methods regardless of task nature, discussed more in \cref{Design Considerations}. These trends align with findings from Bicho et al.~\cite{yu:tel-03313805}, which states that prior user actions fuel the resultant interactions that occur between the user and system.


\subsection{Strategy Analysis}

Our third research question examines the strategies participants adopt when using natural multimodal interfaces for assembly tasks in VR based on the nature of the task. As detailed in \cref{Task Design}, despite being untrained, participants developed specific strategies for task completion, according to our findings.

The first strategy was put-that-there instruction delivery, which aligns with Bolt's theory~\cite{bolt1980put}. As participants were told to interact with the system naturally, they immediately delivered high-level commands (e.g. "pick up the red block") instead of low-level commands (e.g. "move 20 centimeters forward"). This shows user inclination towards intuitive instruction use due to its efficiency and perceived system understanding of their commands. Participants even extended this practice to place multiple objects at once as explained before, with indications being delivered via gestures and gaze.

The second strategy was varying location specification methods. Participants in Task 1 mainly referenced recently placed LEGO as concrete spatial anchors for object placement via description, e.g. "place dark blue on top of the bottom green". In contrast, for Task 2, participants mainly used gestures or gaze for location specification in addition to voice, e.g. "put a resistor right here", due to a lack of concrete spatial anchors on the PCB board. As a result, participants evidently used more unimodal and bimodal commands in Task 1 while they used more trimodal commands in Task 2 (see \cref{section: Modal Analysis}).

One spatiotemporal pattern observed was participants preferring to place objects adjacent to recently placed objects, occurring more in Task 1, which is shown in analyses on the types of utterances used and modal combinations employed (more unimodal, explicit commands were used in Task 1, see \cref{section: Modal Analysis}, \cref{section: Explicit/Implicit Utterances Ratio} respectively). This was more difficult in Task 2 as objects were spaced out across the PCB board, which prompted participants to place the similar objects in their corresponding places, aided by gesture and gaze indications. As such, participants frequently said "[get] another one", with reference to the similar object. 

These strategies collectively imply that while assembly tasks share similar qualities which influence certain behaviors even in users that are untrained, variances in task characteristics drive users towards specific task behaviors. This demonstrates the need for NMUIs to be flexible and capable of interpreting diverse, context-dependent multimodal inputs. As such, these strategies reflect the natural interactions that occur between user and system that would serve as a foundation for NMUIs.

\subsection{Design Considerations} \label{Design Considerations}

\subsubsection{Voice} \label{Design Considerations Voice}

\textbf{Voice primarily initiated interactions}, largely due to its descriptive abilities. Voice input is also uniquely \textbf{able to give temporal references to details} such as previous actions and object properties efficiently. However, its limitations include lack of spatial precision and time to relay information. Voice is \textbf{most effective for conveying descriptions}, especially temporal ones, necessary for understanding and initiating tasks, providing a foundation upon which more precise modalities can build. 

\subsubsection{Hand Tracking} \label{Hand Tracking}
Participants favored hand tracking for \textbf{spatial indications within close to mid-range distances} as tasks progressed. Gestures also \textbf{efficiently convey representational, symbolic, or social cues} such as a thumbs-up to indicate approval. However, its \textbf{limitations include poor performance over long distances and its inability to relay complex instructions} effectively compared to voice. Consequently, \textbf{gestures excel in precise physical manipulation and symbolic/spatial communication}, e.g. demonstrating machinery operation or task completion confirmation.


\subsubsection{Eye Gaze and Head Pose} \label{Eye Gaze and HMD}
Head pose and eye gaze share an integral relationship in VR, as they tend to inherently coordinate their movements for optimal visual perception. Head pose provides broader visual context while eye gaze offers specific points of focus. This coordination is essential for tasks requiring detailed spatial awareness and precise interaction with the VE, improving overall user experience and task performance. \textbf{The combination of head pose and eye gaze excels in relaying where the user's attention lies as well as providing proper spatial indications at all distances}. Its \textbf{limitations include a lack of ability to relay complex instructions} like gestures, as well as some ambiguity if the user speaks without using these modalities to supplement their speech. This may result in misinterpretation of user intent.

\subsection{Generalizability and Transferability}
Our analyses and findings are such that they may be applied to other applications that involve assembly or its attributes, such as location or object selection. For example, in medical simulations where robots with extreme precision can operate on an individual, the user is able to indicate to the robot which instruments to manipulate as well as where to make an incision \cite{reddy2023advancements, iftikhar2024artificial}. In manufacturing, where systems meet AI for intelligent control, location specification for machines and action initiation such as cutting, screwing, etc. are synonymous with the placing and moving commands in terms users instructing simplistically \cite{sahoo2022smart}. Finally, in prototyping applications, behaviors for physical transformations may be derived for actions such as moving objects and performing operations on them \cite{camburn2017design}. It should be noted that while assembly was modeled precisely in these scenarios, the analysis on data performed yields general, higher level behaviors that are expected to be transferred to other similar tasks.

\section{Limitations and Future Work} \label{Limitations and Future Work}
Our elicitation study observed user interactions with the assembly task system to collect a scenario-specific dataset. However, the specific design and prototype of our study introduce limitations to our application.

Our first limitation was in the tasks employed, which were short-ranged assembly tasks that followed Bolt's put-that-there paradigm. As such, more complex processes and operations are unaccounted for, e.g. instructing a CNC machine to execute complex geometrical descriptions for machining or asking the arm to pour a specific amount of liquid into a container. In the future, we could extend the application of the robot to even more domains in terms of simulated HRI in VR.




Our second limitation was the population demographics. The selected population supports the generalization of our results as we investigate high-level behaviors among individuals. Nevertheless, a relatively younger population performed the tasks and provided the data used. In addition, as we did not account for disabled participants, our dataset nor observations do not elaborate findings pertaining to this population. Future work would involve including a wider array of individuals in our study with respect to age, disabilities, experience with certain technologies, etc.



Our third limitation was the amount of interaction modalities employed. We only used natural ones (voice, hand gestures, eye and head gaze) for our analysis, and did not include other such as full body tracking or facial expressions. In addition, traditional methods such as controllers were not used for instruction delivery. Future work could include the use of additional/alternate modalities for analysis to see how these modalities interact with each other for system communication.



Finally, our application for natural multimodal user interfaces was limited to assembly tasks, which does not necessarily infer that our dataset can be expanded to other types of tasks. Our investigation into these types of tasks were meant to provide a use case into the development of a more general multimodal natural user interface. Thus, our investigations in the future will lead us to explore other domains~\cite{gottsacker2025xr}. For example, data visualization, gaming, medicine and pedagogical instances.


Future work aims to develop an autonomous natural user interface by creating a machine learning model that replaces the Wizard-of-Oz method. This model will utilize modern technologies like Large Language Models (LLMs) and Vision Language Models (VLMs) ~\cite{chang2024survey, shafique2025culturally, vayani2025all} to process natural inputs and provide contextual feedback. This advancement could enable users to perform tasks intuitively without prior knowledge or training.


\section{Conclusion} \label{Conclusion}
Our work aims to advance our understanding of users' natural interactions with systems during a VR assembly task without having any prior training or experience. This is done through an analysis of data collected that captures a comprehensive range of multimodal interactions, including voice, eye gaze, hand tracking, and head pose. We collected this dataset through a Wizard-of-Oz experiment, which is intended to provide useful insight into the dynamics of natural modality usage in assembly tasks.

Our findings show participants completing these tasks extend behaviors from the basic form of "put-that-there", with natural modalities having dependence on each other to efficiently complete tasks simulated VR, as well as that modalities have spatiotemporal dependencies on each other. Our findings also indicate that task nature (Collaborative vs. Instructive) has an impact on the types of instructions and modalities participant use to instruct the virtual robot to perform actions, which influence their task strategies. The implications of our work extend to the design of more intuitive and effective NMUIs in simulated manufacturing/operation and assembly tasks in virtual settings, with use cases including rapid prototyping, training/learning, and robotic surgery. Such interfaces could be tailored to understand and predict human intentions more accurately in the future, which can bring about an increase in efficiency and the minimization of human and system error. Moreover, the proposed methodology in our study can serve as a framework for future work aimed to explore complex interaction scenarios in VEs. Through our dataset and findings, researchers can better understand the intricacies of natural instructions provided to NMUIs which ultimately lead to the development of more robust and adaptive interfaces of this kind. In addition, this work serves as a stepping stone into the exploration of how users interact with arbitrary systems across multiple domains.

\bibliographystyle{abbrv-doi}

\bibliography{template}
\end{document}